# Experimental Measurement of Overlapped Sheaths

Mudi Chen, Michael Dropmann, Ke Qiao, Zhiyue Ding, Lorin S. Matthews and Truell W. Hyde

*Center for Astrophysics, Space Physics and Engineering Research (CASPER), One Bear Place 97283, Waco, Texas, Baylor University, 76798-7316, USA*

**Abstract:**

Due to the complicated environment of the plasma sheath, it is difficult to be experimentally measure plasma characteristics in a narrow geometry where the sheaths from opposite boundaries overlap. Such geometries are often found in industrial plasma applications. Here we employ micron-sized dust grains as non-perturbative probes of the plasma environment. A particle-freefall technique is used to measure the sheath profiles produced by a rf plasma within a glass box. The results show that this technique can identify the plasma operating conditions for which the sheaths on opposite walls begin to overlap as well as the magnitude of the effect.

# I Introduction:

Materials immersed in a low temperature, partially ionized plasma become negatively charged due to velocity differences between the electrons and more massive ions. Once charge equilibrium has been reached, electrons are repelled from and ions are accelerated toward the material surface, creating a spatial region where the electron and ion densities are not equal: the plasma sheath. The potential difference between the material surface and the quasi-neutral bulk plasma gives rise to an electric field which accelerates ions from the plasma bulk to the surface. The sheath width is often defined as the distance from the surface to the point at which the ions have been accelerated to the ion acoustic speed $c_s = \sqrt{kT_e/m_i}$, a condition known as the Bohm Criterion [1]. This point is also sometimes referred to as the sheath edge [2]. Since the Bohm speed is much larger than the ion thermal speed in an rf plasma due to the fact that $T_e \gg T_i$, the ions require an electric field to accelerate them to the Bohm speed. The region in which the electric field is larger than zero and the speed of ion stream is less than Bohm speed is defined as presheath [1]. Therefore, the region of space influenced by the plasma facing surface includes the sheath and the presheath, which in this paper we shall call the sheath width $s_0$.

It is well known that experimental data defining the properties of the sheath for even simple geometries, such as planes, cylinders and spheres, is difficult to obtain due to the fact that the sheath is often narrow and easily perturbed. A number of diagnostic methods have been used to circumvent this problem, including emissive Langmuir probes [3-4], laser-induced fluorescence [5] and charged microspheres (i.e. dust particles) acting as probes [6-12] with each bringing its own challenges. Langmuir probes are difficult to use to measure the sheath characteristics as they alter the sheath environment. Optical methods, while non-perturbative, require both laser accessible energy-level transitions and long averaging times in order to acquire reliable data. Using dust particles as probes, although difficult, is considered as one of the better ways to diagnose the sheath environment due to their small perturbation and rapid response. An early application of this method can be found in Samarian et al. [7,10,13], where floating dust particles were used to measure the overlapped sheath above a trench by assuming the observed floating position of the dust particles marked the sheath edge.

A proper understanding of the physics underling the production of the plasma sheath is vital to the plasma processing industry. Processes such as plasma enhanced deposition, etching, and plasma immersion ion implantation, are conducted in the plasma sheath, and the sheath electric field determines the ion densities and dynamics. The target geometry is often complicated, with a trench-like geometry being the most widely used [14-16]. The sheath produced by a trench geometry will in general fit one of three different types depending on the sheath width in relation to the trench width W and depth D. As shown in Fig. 1a, a sheath width much smaller than the width of the trench $s_0 \ll W$ allows the plasma edge to conform to the shape of the surface topography, a feature known as plasma molding [17]. As shown, there is no overlap between the sheaths formed by the different surfaces of the trench. When the sheath width is comparable to the width $s_0 \approx W$, the sheaths produced at the walls of the trench begin to overlap (Fig. 1b). Compared with case (a), the edge of the sheath is raised from the trench floor or even pushed out of the trench. When the sheath width $s_0$ is much larger than the width $s_0 \gg W$, the sheaths of the two opposing walls completely overlap (Fig. 1c). The sheath edge is almost planar, as it would be if there were no trench.

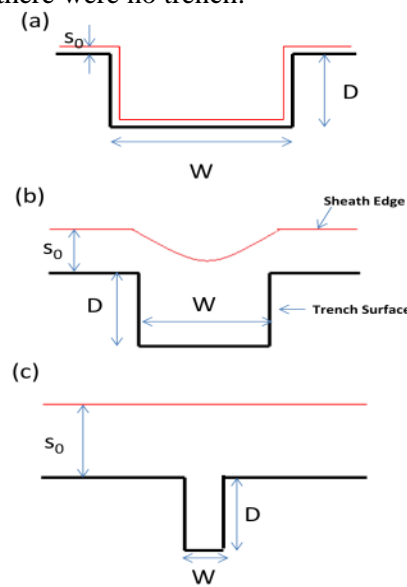

Fig.1. Schematic of a typical plasma sheath produced by a trench with width W and a depth D. The geometry of the sheath boundary, indicated by the red line, depends on the sheath width $s_0$ in relation to the trench width, W.

Experiments mapping the sheath profile produced inside a trench and for overlapping sheaths are rare. In an experiment reported by Sheridan [26], the overlapping is considered to be due to the sheath of the top surfaces of the opposite trench steps. To data, there have been no experimental measurements inside a trench. Recently, Chen et al. reported a free-fall technique [18], which allows measurement of the sheath profile through analysis of the trajectories of freely falling dust particles. This technique produces very small perturbations to the plasma and allows mapping of the electric force within the sheath. In the experiments reported here, the particle free-fall technique is used to map the sheath electric field in the vertical direction as a function of the changing sheath width, which is altered by varying the plasma operating power and pressure. A glass box (consisting of insulating walls and a conductive substrate or electrode) is employed to simulate a trench geometry. The experimental setup and method are described in Sec. II, experimental results are given in Sec. III, and conclusions are presented in Sec. IV.

## II Experimental Setup and Method

The experiment was performed in a modified gaseous electronics conference (GEC) radio-frequency (RF) cell, filled with argon. The gas pressure was varied from 60 mTorr to 140 mTorr and a RF electrical field was produced by a pair of capacitively-coupled electrodes 8 cm in diameter, separated by a distance of 2.74 cm. The upper electrode was grounded, while the lower electrode was powered by a RF generator at a constant frequency of 13.56 MHz. The amplitude of the input RF power was varied between 1.70 - 10.71 W. An open-ended glass box of dimension 2.54 cm $\times$ 2.34 cm $\times$ 2.34 cm (height $\times$ width $\times$ length) and 2 mm wall thickness was placed at the center of the lower electrode. A dust dropper was employed to introduce particles (melamine formaldehyde spheres with manufacturer-specified mass density 1.514 g/cm$^3$ and diameter 8.89 $\pm$ 0.09 μm) into the glass box. The particles were illuminated by a vertical sheet of laser light, and their positions were recorded at 1000 frames per second (fps) using a side-mounted, high-speed CCD (Photron) camera and a microscope lens.

The sheath within a glass box is much more complicated than that above the planar electrode. The sheaths for each wall of the glass box might interact with one another, which is similar to the case in the trench-like geometry mentioned above. Several experimental measurements in the overlapping sheath region have been reported [13, 27], but most of them can only measure the place where the dust particles is levitating. Here, the particle-freefall method was first applied to map the electric field force profiles in the sheath of the planar lower electrode under different gas pressures and rf powers. Then, the electric field force profile inside the glass box was investigated for the same set of operating conditions.

In the particle-freefall measurement, a few hundred dust particles are dropped from the dust dropper into the glass box and their trajectories are recorded by the high speed camera. A MATLAB-based algorithm [19] for particle detection was employed to identify the particles in each recorded frame, which were then linked using the Hungarian algorithm [20]. Resulting particle trajectories with minimal particle movement or discontinuities in the calculated accelerations were eliminated, as these are usually due to noise falsely interpreted as particles. Particle-particle interactions were managed by establishing a minimum allowable distance between detected particles, in this case, 0.5 mm. The electric force $Q_d E$ acting on each particle was then calculated from the acceleration, assuming that the only other forces acting on the particles were gravity and neutral gas drag [18]. The equations of motion in the vertical and horizontal directions are then given by

$$m_d \ddot{Z} = -\beta \dot{Z} + Q_d E_z - m_d g \qquad (1)$$
$$m_d \ddot{X} = -\beta \dot{X} + Q_d E_x \qquad (2)$$

where $m_d$ is the mass of the dust particle, $\beta$ is the gas drag coefficient, g is the acceleration due to gravity (9.81 m/s²), $Q_d$ is the charge on the dust particle, and $E_z$ and $E_x$ are the electric fields in the vertical and horizontal directions. In this case, the drag is considered to be due to collisions with the neutral gas particles. The ion drag force is estimated to be more than an order of magnitude smaller than the confinement forces, and is neglected here. The details of this method and the force analysis can be found in our previous work [18]. With the accelerations of the particles calculated from the trajectories, the confining forces $Q_d E_z$ and $Q_d E_x$ can be calculated.

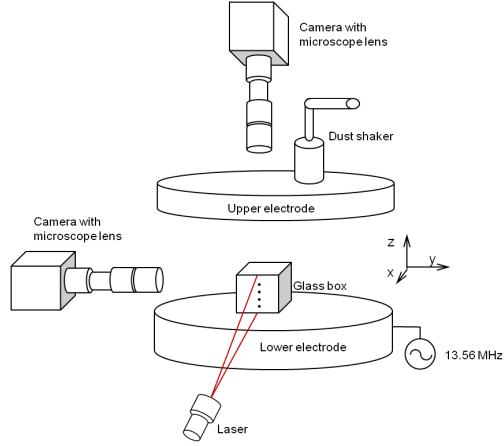

**Fig. 2. Experimental setup. The upper ring electrode is grounded while the lower electrode is powered. T The open-ended glass box shown has outer dimensions of 2.54 cm × 2.34 cm × 2.34cm (height, width, length) is placed on the lower powered electrode. The motion of the particles is captured using a Photron CCD high speed camera (side view).**

## III Experimental Results

Experimentally measured results for the vertical electric field force profiles in the sheath of the planar lower electrode are shown in Fig. 3, which compares the effects of changing rf power (Fig. 3a) and changing gas pressure (Fig. 3b). Note that all forces here and elsewhere in this paper are normalized by the gravitational force acting on the dust particle.

As a dust particle enters the sheath from the plasma bulk above, the vertical electric force rapidly increases from zero as the magnitude of the electric field increases. The gradient of the force curve decreases as the particle approaches the lower electrode. This is due to the fact that the electron density drops quickly near the lower electrode, leading to a reduced magnitude of charge on any individual dust particle [21].

In the plasma bulk, the electrostatic force acting on the dust particles should be almost zero, the result shown in Fig. 3 can therefore be used to establish the sheath width $s_0$, the distance above the lower electrode separating the sheath and presheath from the plasma bulk [28]. Here, we define the sheath width as the first point above the lower electrode (Z=0), where the force curve becomes zero. It should be noticed that the electric field force in the region above the cross point is not complete zero, but this is inside the error bars of this measurement technique, which is about ±0.05 $m_d g$.

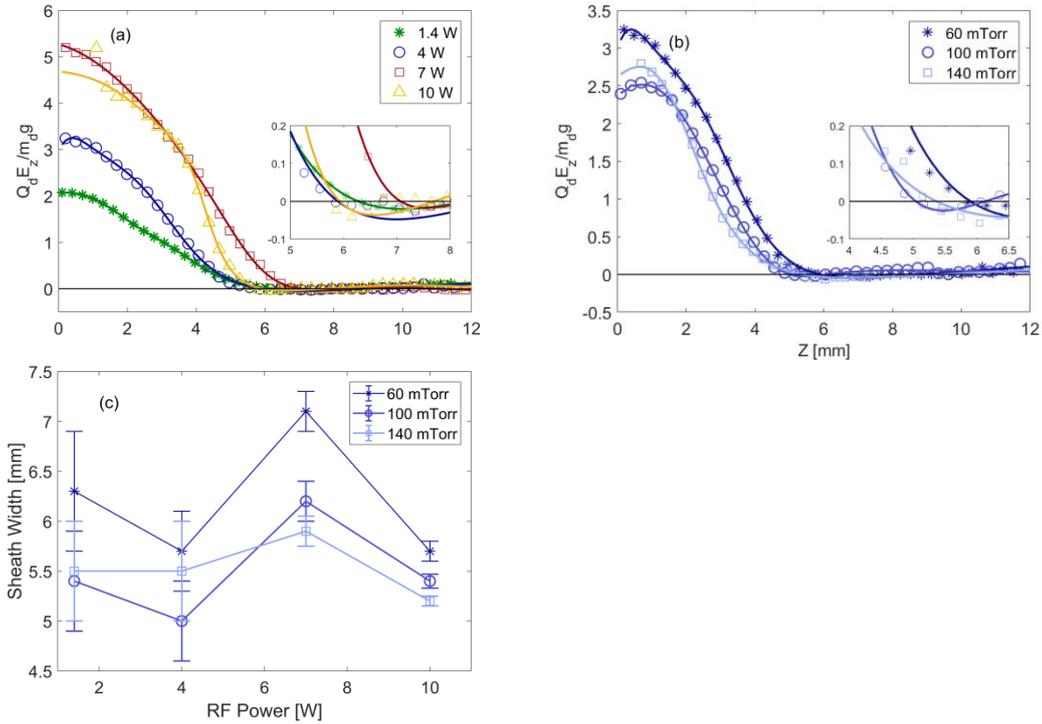

Fig. 3. Vertical electrostatic force acting on dust particles levitated in the sheath above the lower powered electrode as a function of distance (Z) from the lower electrode for (a) varying rf powers and fixed gas pressure of 60 mTorr and (b) varying gas pressures for a fixed rf power of 4 W. The symbols indicate measured data points while the lines indicate a rational fit. The inserts show a detailed view of the region where the vertical force goes to zero, marking the sheath width. (c) Sheath widths determined for different powers and gas pressures.

It can be seen from Fig. 3(c) that the sheath width is only weakly related to the plasma power. The largest sheath width observed across all gas pressures occurred at 7 W, although the differences between the sheath widths measured at all other powers fell within the error bars. The reason why this particular sheath exhibited a different sheath width at a power of 7 W is still unknown. The sheath width was observed to be largest at a gas pressure of 60 mTorr, although no significant differences between sheath widths was observed for pressures of 100 mTorr and 140 mTorr.

The measured electric force at the center line of the glass box is compared with the electric force measured above a planar lower electrode in Fig. 4. At the lowest gas pressure P = 60 mTorr, the vertical electric force profile within the box can be seen to be quite different from that measured within the sheath above the planar electrode. The profiles of the vertical electric force above the planar electrode (Fig. 3 (a)), indicated by dashed lines in Fig. 4 (a), show that the electric force acting on the dust particles continuously increases as the particles approach the lower electrode, flattening out very close the electrode surface (usually less than 0.5 mm above the lower electrode). Inside the box, the point marking the sheath width, where the vertical electric field force is approximately equal to zero, is much further from the lower electrode than without the box. Overall a large vertical extent, the gradient of the electric force is relatively small, or even becomes negative. This leads to a relatively 'flat' region indicated by the shaded region in the electric force profile as shown in Fig. 4(a), which is not observed in the planar sheath cases. The extent of the flat region decreases with increasing rf power. At the lowest operating power (1.4 W) the length of this flat region is much larger than the sheath width measured for the planar electrode. At fixed operating powers, as shown in Fig. 4(b), increasing the gas pressure leads to a convergence between the force profiles observed within the box and those observed above the lower planar electrode. The 'flat' region disappeared at 100 and 140 mTorr gas pressure.

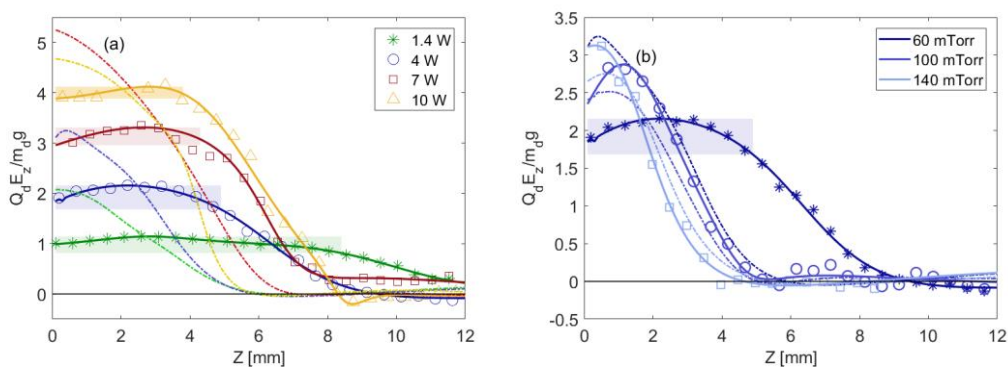

**Fig.4. Comparison of vertical electric field force in the sheath of the lower planar electrode and inside the box for (a) varying rf powers at a fixed gas pressure of 60 mTorr and (b) varying gas pressures for a fixed rf power of 4 W. Symbols show the measured data points, solid line indicate the corresponding fit and the dash line show the fits from the data shown in Fig. 3.**

Fig. 5 (a) shows a 2-D map of the horizontal electric field force for the lowest pressure and power (60 mTorr gas pressure and 1.4 W rf power). Fig 5(d) shows the 1-D profile of the horizontal electric force at the vertical position marked by the red line in Fig. 5(a). At this height, the horizontal electric force is almost linear across the region [-7 mm to 7 mm], indicating that the horizontal force from the walls is felt across the entire region and the sheaths of the glass walls overlap with one another. The map shown in Fig. 5(b) corresponds to the plasma conditions for the case where there is a small flat vertical force profile. At positions near the lower electrode, indicated by the red line in Fig. 5(b), the horizontal electric force is nearly linear across the width of the box, as shown by the corresponding 1-D profile in Fig. 5(e). This leads to the conclusion that the sheaths from the walls are overlapping in this region. At distances farther from the lower electrode, as indicated by the blue horizontal line in Fig. 5(b), the corresponding 1-D profile is nearly flat cross the center of the box, increasing rapidly in magnitude only very close to the walls. This shows that the sheaths from the vertical walls are well-separated at this height. Finally, a horizontal electric force map is shown in Fig. 5(c) for an operating condition where the vertical electric force profile within the box is similar to that above the planar lower electrode. The horizontal electric field is nearly zero throughout the entire box region, only having significant non-zero magnitude in the corners of the box very close to the lower electrode surface.

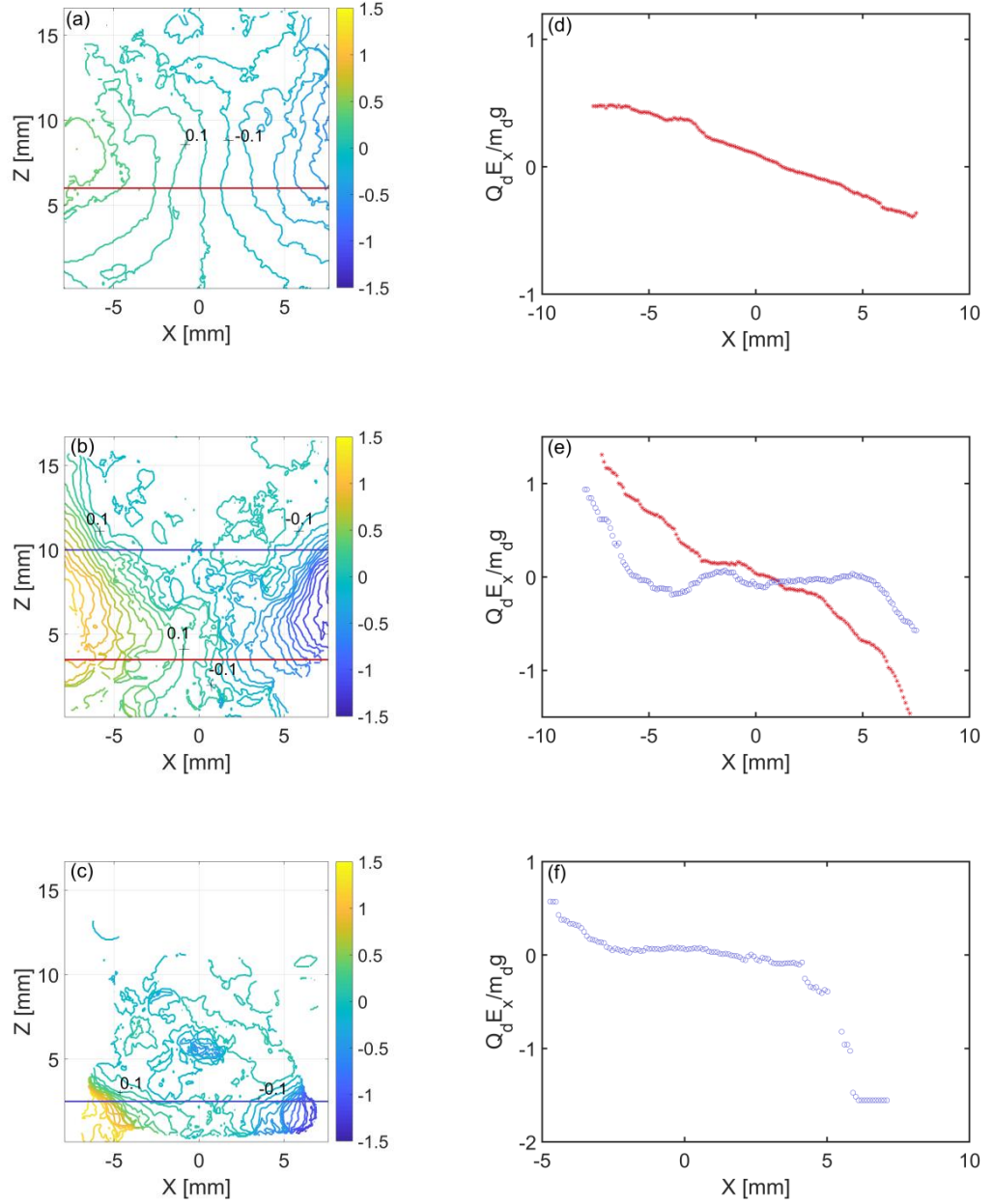

Fig.5. The horizontal electric field force acting on the dust particles within the box at (a) 60 mTorr gas pressure and 1.4 W rf power; (b) 60 mTorr gas pressure and 4 W rf power and (c) 140 mTorr gas pressure and 4 W rf power. The colorbar indicates the magnitude of the electrostatic force normalized by the force of gravity, with positive forces acting to the right. (d-f) Variation of horizontal electric force at the vertical positions marked by the horizontal lines in (a-c), respectively.

# IV Discussion & Conclusion

The region where overlapping between the sheaths of the inner surface of the glass box can be identified from the data shown in Figure 5. The sheath overlap is indicated by the differences between the profiles of the horizontal electric field in the regions marked by the blue and red lines in the force maps. In the region where the sheaths overlap, the contour line indicating $\frac{Q_d E_x}{m_d g} = 0$ exists only in the middle of the box (X = 0) with the contour lines marking $\frac{Q_d E_x}{m_d g} = \pm 0.1$ diverging from one another when approaching the lower electrode, as shown in Fig. 5(a) and for the region Z < 6.5 mm in Fig. 5(b). In a region where the sheaths are not overlapped, the contour lines denoting $\frac{Q_d E_x}{m_d g} = \pm 0.1$ converge with decreasing Z, and $\frac{Q_d E_x}{m_d g} = 0$ over a large range inside the box (the regions ( -1mm < X < 1 mm or 5mm < Z in Fig. 5 (c) and Z > 6.5 mm in Fig. 5(b)).

In the experiments presented here, the overlapping is observed between the sheaths of the inner surfaces of the glass walls, instead of the top surfaces of the glass walls stated by Sheridan [26]. In some cases, such as that shown in Fig. 5(b), there is no sheath overlap in the upper region of the box, while the sheath overlap in the lower part of the glass box.

The transition from non-overlapping to overlapping sheaths from the glass walls is most likely due to the decreased density of electrons and ions. Sheath theory indicates that the closer to the lower electrode surface, the smaller the electron and ion densities [28] and this depletion may be greater within the box due to the absorption of electrons on the glass walls. Decreasing the electron and ion densities leads to a larger Debye shielding length, which in further means the sheath influence extends further, increasing the sheath width. Given that the glass box is immersed in the sheath of the lower electrode, an increasing shielding length for glass wall can be observed when approaching the lower electrode. As this sheath width increasing to larger than the half of the width of the glass box, the sheaths of the glass walls begin to overlap.

Comparing the 'flat' regions for the vertical electrostatic force profiles (Fig. 4) and the overlapping regions in the 2D maps of the horizontal electrostatic force (Fig 5) shows that the 'flat' region and the overlapping sheath region coincide with one another. For example, in the case with the greatest sheath overlap (60 mTorr gas pressure and 1.4 W rf power), the 'flat' region shown in Fig. 4(a) extends from approximately zero to 8.3 mm in Z, while the sheath overlap can be observed in almost the same region in Fig.5 (a). For the case with the same pressure but at an increased rf power of 4 W, the 'flat' region in Fig. 4 (a) extends from Z = 0 to Z = 6 mm with the overlapping sheath region in Fig. 5(b) at approximately Z = 0 to Z = 6.5 mm. At higher gas pressures, no overlap is evident in the map of the horizontal electrostatic force shown in Fig. 5(c) and the corresponding vertical electric field force profiles in Fig. 4(b) exhibit no 'flat' region.. Therefore, the 'flat' region in the vertical force profiles can only be observed in a region where sheath overlap exists.

In summary, the particle-freefall technique has been used to measure the sheath profile both for a planar electrode and inside a glass box. This method has been shown to accurately measure the electrostatic force in the sheath and the sheath width for various power and pressure settings. The force profiles clearly identify conditions which produce overlapping sheaths from the inner surfaces of the glass box, as would expected inside a trench. The overlap leads to a linear electric field force distribution in the horizontal direction and a 'flat' curve in the vertical electric field force profile.

## Acknowledgement

Support from NSF grants PHY-1740203 and PHY-1707215 and NASA / JPL Contract 1571701 is gratefully acknowledged.